# HIR: A collisional-radiative model for hydrogen ionization in plasma with anisotropic fast ions

S.V.Polosatkin

Budker Institute of Nuclear Physics, Novosibirsk, Russian Federation
Novosibirsk State University, Novosibirsk, Russian Federation
Novosibirsk State Technical University, Novosibirsk, Russian Federation
*s.v.polosatkin@inp.nsk.su*

The paper presents the HIR (Hydrogen Ionization Rates) software package developed for calculating the level populations of atomic hydrogen and the dynamics of its ionization in hot plasma, including the case of fast ions with an anisotropic distribution function. The code is based on a steady-state collisional-radiative model that includes electron and ion impact excitation, de-excitation, ionization, charge exchange, and spontaneous radiative transitions. It is demonstrated that stepwise ionization via excited states contributes up to 20% to the total ionization rate and requires including levels up to the principal quantum number N=8. For the GDML and GOL-NB mirror trap facilities, the effects of stepwise ionization on the mean free path of fast atoms are analyzed. It is shown that collisions with different plasma species are non-additive due to stepwise ionization, leading to a reduction in the mean free path by up to 7% at 40 keV. The anisotropy of the fast ion distribution is found to have a negligible effect (less than 1%) on the ionization dynamics in the GDML plasma. The ionization cost of hydrogen atoms in the GOL-NB plasma is calculated, demonstrating a strong dependence on plasma density due to competition between collisional and radiative de-excitation. The HIR package is openly available on GitHub.

## Introduction

In magnetic confinement fusion plasma facilities, the interaction of hot plasma with hydrogen (or its isotopes) plays an important role. Neutral hydrogen can enter the plasma from the region between the plasma boundary and the vacuum chamber wall (via plasma fueling or wall desorption), be formed on plasma-facing components (through neutralization of incident ion fluxes), or be injected as fast neutral beams for heating or diagnostics. At plasma temperatures above a few electron-volts, ionization processes dominate over the reverse recombination processes, and the latter can be neglected, i.e., a neutral atom entering the plasma loses an electron after some time and becomes an ion. The dynamics of hydrogen atom ionization determines the depth of their penetration into the plasma and the ionization cost (i.e., the energy expended on ionizing an atom, taking into account the line radiation accompanying ionization). It is well known that for hydrogen atoms in plasma, stepwise ionization, i.e., excitation of atoms to one of the upper levels followed by ionization from that level, plays a significant role. With regard to fast neutral beams in plasma, this effect has been examined in detail and described in the classic work [1], in which attenuation coefficients for neutral beams with energies from 100 to $10^4$ keV in fusion plasma with temperatures in the range of 1–50 keV are calculated.

One of the research directions in the field of nuclear fusion is the study of plasma confinement in linear magnetic systems (mirror traps) [2]. In such systems, the ion distribution function is non-Maxwellian and anisotropic due to the presence of a loss cone in velocity space. This effect is most pronounced when fast neutral injection is used for plasma heating. In particular, in the project of the GDML facility currently being developed at the Budker Institute of Nuclear Physics [3], it is assumed that the plasma contains two ion populations with comparable densities – a "warm" one with an isotropic Maxwellian distribution function and a temperature of about 1 keV, and a "fast" one, significantly anisotropic, formed during the slowing down in the plasma of ions injected into the plasma using a neutral injection system. The case of an anisotropic ion distribution function in the plasma is not considered in [1], which motivated the development of the HIR (hydrogen ionization rates) software package for calculating the level populations of atomic hydrogen and ionization dynamics in plasma. This paper presents a description of the physical models and atomic data used in the package, as well as estimates of the effect of stepwise ionization and anisotropy of the ion plasma component on the ionization dynamics in plasma. The software package itself is openly available ([4]).

## 1. Level population model

To describe the level populations and ionization dynamics, a steady-state collisional-radiative model is used. Steadiness here means that the characteristic times of changes in external model parameters (such as plasma temperature and density) are longer than the lifetimes of the excited states of the atom entering the plasma. The model accounts for excitation, de-excitation, and ionization in collisions of atoms with plasma electrons and ions, as well as spontaneous radiative transitions between levels and charge exchange (electron exchange between an ion and an atom). The sources of atomic data are discussed in Section 2.

To determine the level populations of the hydrogen atom, a system of equations is solved for the concentrations of hydrogen atoms at levels with different principal quantum numbers:

$$\sum_{k=i+1}^{n_{max}} n_k A_{ik} - \sum_{k=1}^{i-1} n_i A_{ki} - n_i\left(\nu_i^{ion_e} + \nu_i^{ion_i} + \nu_i^{CX}\right) + \sum_{k=i+1}^{n_{max}} n_k\left(\nu_{ik}^{deex_e} + \nu_{ik}^{deex_i}\right) - \sum_{k=1}^{i-1} n_i\left(\nu_{ki}^{deex_e} + \nu_{ki}^{deex_i}\right) = \Gamma_i \quad , \quad i = 1,2,3 \ldots, i_{max} \quad (1)$$

Here $\nu = n_{pl}\langle\sigma v\rangle$ is the rate coefficient for a given process, $n_{pl}$ – is the density of the plasma species involved, $A_{ki}$ – are the spontaneous transition probabilities, and $\Gamma_i$ – incoming flux of atoms with an electron on level i.

The atomic level population $n_i$ can be used to calculate quantities characterizing the interaction of the atom with different plasma components. For example, the global lifetime of atoms prior to their conversion to ions (due to ionization or charge exchange) is:

$$\tau = \frac{\sum_i n_i}{\sum_i \Gamma_i}$$

The partial lifetime prior to a specific process is:

$$\tau_p = \frac{\sum_i n_i}{\sum_i n_i \cdot \nu_i^{spec}}$$

## 2. Atomic data for calculations

Data on the cross sections for ionization, excitation, and charge exchange were taken from the work [5]. These cross-section data can also be found in the available online atomic cross-section databases Aladdin [6] and CollisionDB [7]. We draw the reader's attention to several inaccuracies in these sources that may lead to errors in determining the cross sections:

• In the report [5], the formula for the ionization cross section from levels with n > 3 (process 1.2.6) is missing a factor of $n^2$.

• In the formulas for excitation and ionization cross sections in the report [5], the upper level is denoted by the letter m and the lower level by n, but in the formula for the oscillator strength in the appendix to the report, the opposite notation is used (n – upper level, m – lower level).

• In the Aladdin database [6], for the charge-exchange cross section from excited levels (n > 1), the value of the normalized cross section ($\sigma/n^4$) is given without any indication that the cross section must be multiplied by $n^4$.

The CollisionDB database contains Python functions for calculating cross sections using the formulas from the report [5], but in our opinion, they are organized in a rather inconvenient manner, as they require inputting individual approximation coefficients for each cross section. Therefore, in the HIR package [4], general functions are used for calculating cross sections, requiring only the velocity of the incident particle and the principal quantum numbers of the upper and lower levels of the transitions, within which the necessary approximation coefficients are computed.

The cross sections for collisional de-excitation can be obtained from the known excitation cross sections using the principle of detailed balance for processes in a thermodynamically equilibrium plasma with a Maxwellian velocity distribution. For collisions with electrons, the relationship between the excitation and de-excitation cross sections can be expressed by the Klein–Rosseland relation [8]):

$$\sigma_{deex}(E) = \sigma_{ex}(E + \Delta E_{ik}) \cdot \frac{g_i}{g_k} \cdot \frac{E + \Delta E_{ik}}{E} \qquad (2)$$

where i,k are lower and upper level numbers of the transition, $\Delta E_{ik}$ – the energy difference between the levels.

For collisions between atoms and protons, the correct application of the detailed balance principle generally requires taking into account the kinetic energy transferred to the atom as a result of the collision. However, this energy is significantly smaller than the initial energy of the ions participating in the collision (see, for example, [9–10]); therefore, we also used formula (2) to calculate the de-excitation cross sections for collisions with protons. A similar approach was used, for example, in the work [11] (formula (5.15) in that work).

The values of the transition probabilities $A_{ki}$ were calculated using the Johnson formula [12].

To verify the correct operation of the code and the accuracy of the elementary process rate data, calculations of the effective ionization rate of slow hydrogen atoms as a function of plasma temperature and density were performed. These calculations accounted only for electron collisions; the electron distribution in the plasma was assumed to be Maxwellian, and the ionization rate was calculated as:

$$s_{eff} = \frac{\sum_i n_i \cdot \nu_i^{ion_e}}{n_{pl} \sum_i n_i} \qquad (3)$$

The obtained effective ionization rates were compared with the effective ionization rate data presented in the openly accessible OPEN ADAS library [13]. This library contains several data sets (files scd89_h.dat, scd96_h.dat, and scd12_h.dat). The number in the file name indicates the year of data publication. The ratios of the effective reaction rates given in the OPEN ADAS database to those calculated using the HIR package are shown in Figure 1.

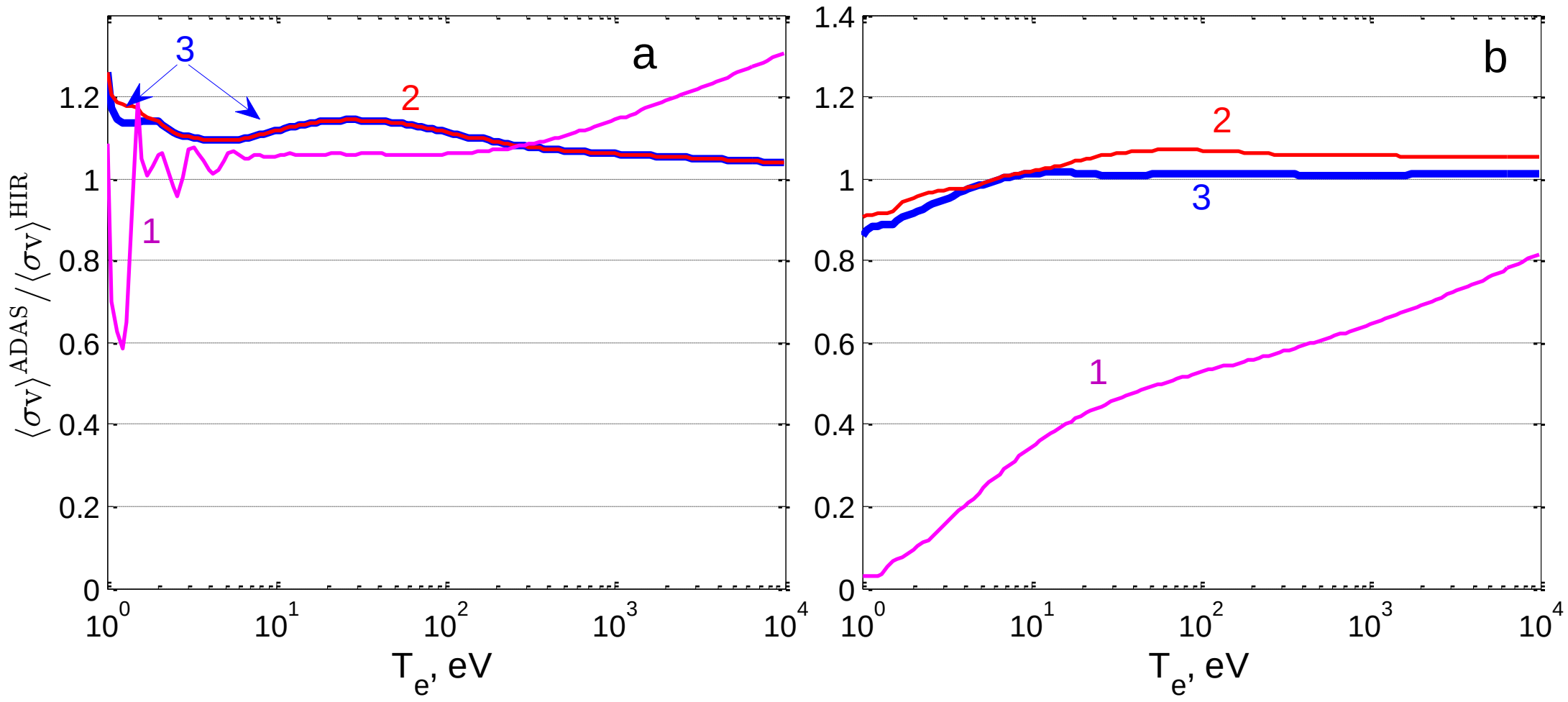

Figure1. Ratio of the effective ionization rates given in the OPEN ADAS database to those calculated using the HIR package. (a) – plasma density of $10^8$ cm$^{-3}$, (b) – plasma density of $10^{15}$ cm$^{-3}$, 1 – scd89_h, 2 – scd96_h, 3 – scd12_h.

It can be seen that for the scd96_h and scd12_h data sets, the deviation from the calculations in the HIR package is about 10%. At the same time, for low plasma density, the contribution of stepwise ionization is very small (about 1%), and the effective ionization rate is determined exclusively by the ionization cross section of the hydrogen atom from the ground state. Therefore, it can be assumed that the difference in the obtained ionization rate values is determined by the difference in the underlying cross-section data used for the elementary processes. Unfortunately, information on the cross sections used in calculating the ionization rates in OPEN ADAS is not publicly available; however, the above comparison serves to assess the accuracy of the [5] cross-section data — data that, in addition to being used in the HIR package, are also commonly utilized in modeling efforts because of their ready availability.

## 3. Stepwise ionization in electron collisions

To assess the effect of stepwise ionization, calculations of the effective ionization rate of slow hydrogen atoms in plasma were performed. As in Section 2, the calculations accounted only for electron collisions, which corresponds to a plasma with a low ion temperature. The effective ionization rate was calculated using formula (3). Figure 2 shows the ratio of the ionization rate from the ground level to the effective ionization rate (which includes stepwise ionization via excited levels). As expected, the stepwise ionization effect is most pronounced at low temperatures and high plasma densities; however, at plasma densities above $10^{13}$ cm$^{-3}$, it becomes significant even in high-temperature plasma.

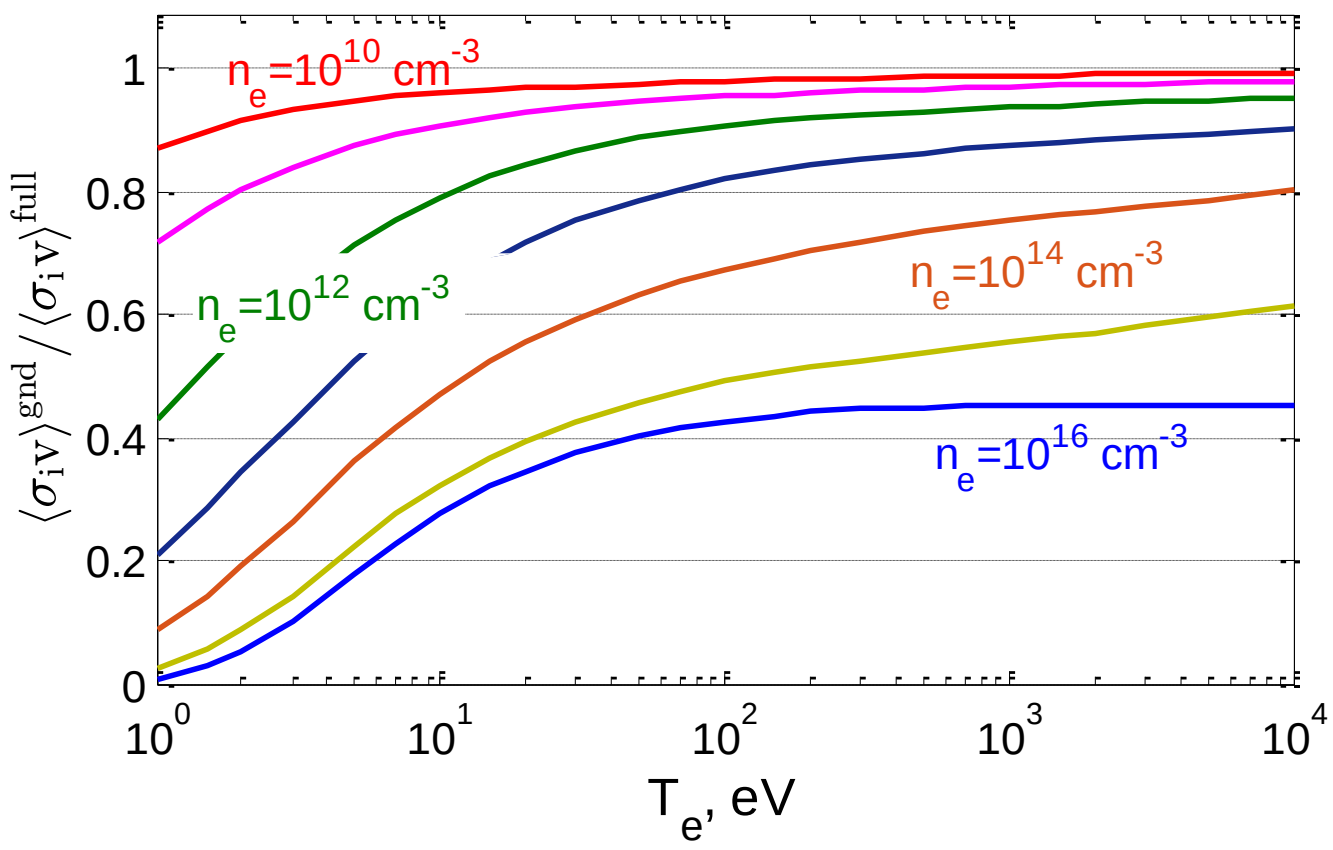

Figure 2. ratio of the ground-level ionization rate to the effective ionization rate.

Figure 3 shows the dependence of the calculated effective ionization rate on the number of levels included in the model at a plasma density of $10^{14}$ cm$^{-3}$ and various temperatures. As can be seen from this figure, levels up to N = 8 make a noticeable contribution to stepwise ionization.

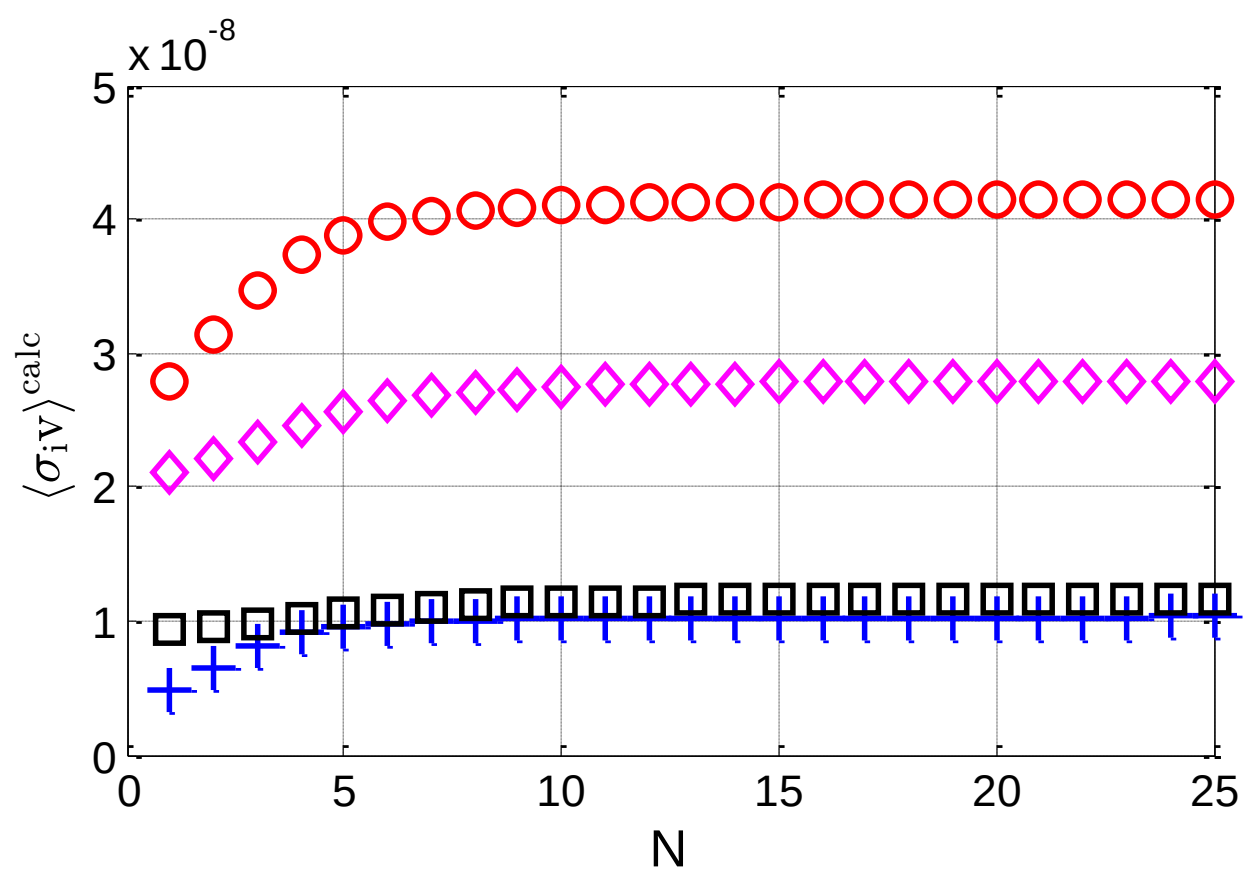

Figure 3. Calculated effective ionization rate versus the number of levels included in the model for different electron temperatures: crosses – 10 eV, circles – 100 eV, diamonds – 1 keV, squares – 10 keV.

## 4. Stepwise ionization of fast atoms in plasma

In open traps with plasma heating by injection of fast hydrogen atoms, an important parameter is the mean free path of a fast atom in the plasma before its capture, which can be related to ionization processes in collisions with plasma electrons and ions, as well as to charge exchange on plasma ions. As an example, we consider here the capture of injected fast neutrals into the plasma in the GOL-NB facility [15] currently operating at the Budker Institute of Nuclear Physics SB RAS and in the designed GDML facility [3].

The GDML facility is a trap with gas-dynamic plasma confinement and plasma heating by means of fast neutral injection. A distinctive feature of this type of facility is the two-component ion distribution function, consisting of a "warm" collisional component, confined gas-dynamically, and a "fast" collisionless component. The warm component has an isotropic Maxwellian distribution, while the fast component is significantly non-Maxwellian and anisotropic.

As a characteristic operational regime of the GDML facility, we consider a regime with electron temperature $T_e$ = 650 eV, warm ion component temperature $T_{iw}$ = 1080 eV, and densities of electrons, warm ions, and fast ions $n_e = 4.5\cdot10^{13}$ cm$^{-3}$, $n_{iw} = 2.8\cdot10^{13}$ cm$^{-3}$, $n_{if}$ = $1.7\cdot10^{13}$ cm$^{-3}$. The distribution function of fast ions at the injection point is shown in Figure 4. The injection energy is $E_0$ = 40 keV, and the injection angle (the angle between the fast atom velocity vector and the facility axis) is 50.9°.

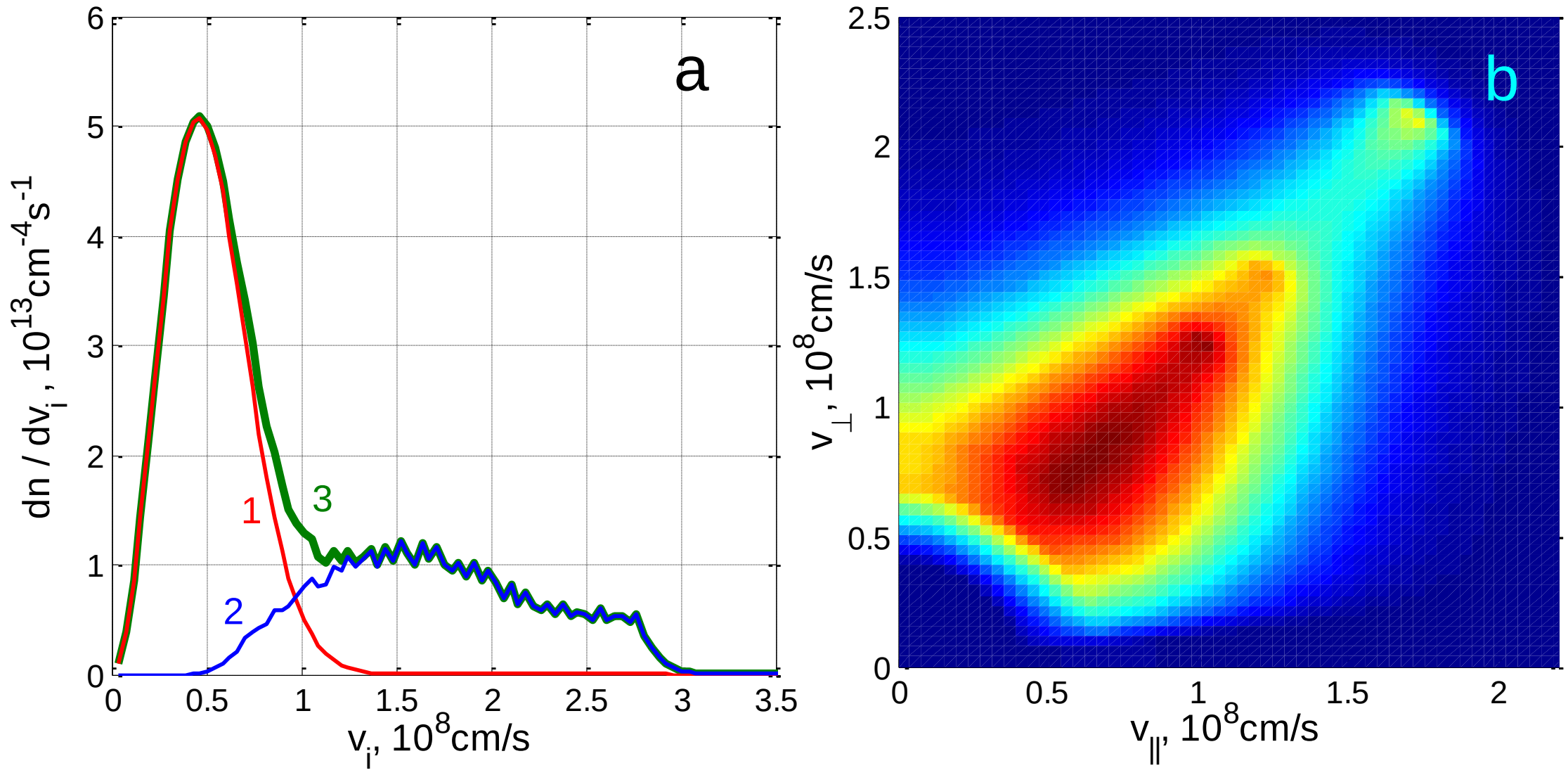

Figure 4. Ion distribution function in the GDML facility. (a) – ion distribution function versus total velocity; 1 – warm component ($T_{iw}$=1080 eV), 2 – fast component, 3 – total distribution function. (b) – distribution of fast ions versus parallel and perpendicular velocities.

The GOL-NB facility, designed to test the efficiency of multi-mirror plasma confinement, has a similar design but substantially lower plasma parameters. For the GOL-NB facility, we consider a characteristic regime with the following parameters: plasma temperature $T_e = T_i$ = 20 eV, electron density $2{\cdot}10^{13}$ cm$^{-3}$, and injected atom energy $E_0$ = 24 keV. The density of the fast ion population in this facility is estimated to be on the order of $10^{11}$ cm$^{-3}$ and does not affect the capture of atomic beams.

Note that for both of these facilities, the fast atom beams injected into the plasma contain not only the main component (atoms with full energy $E_0$) but also atoms with fractional energies ($E_0$/2, $E_0$/3, $E_0$/18).

The dependence of the calculated mean free path of fast atoms at the injection energy on the number of levels included is shown in Fig. 5. It can be seen that stepwise ionization reduces the mean free path by approximately 20%, and, as in the case of electron-impact ionization of slow atoms, levels up to N = 8 make a noticeable contribution to stepwise ionization. Note that for fast ions moving in a magnetic field, there exists an effect of autoionization from upper levels due to the electric field arising in the accompanying reference frame [1]. The function for calculating the maximum upper level is included in the HIR package; for the GDML and GOL-NB facilities, the maximum upper level values are $i_{max}$=11 and $i_{max}$=13, respectively.

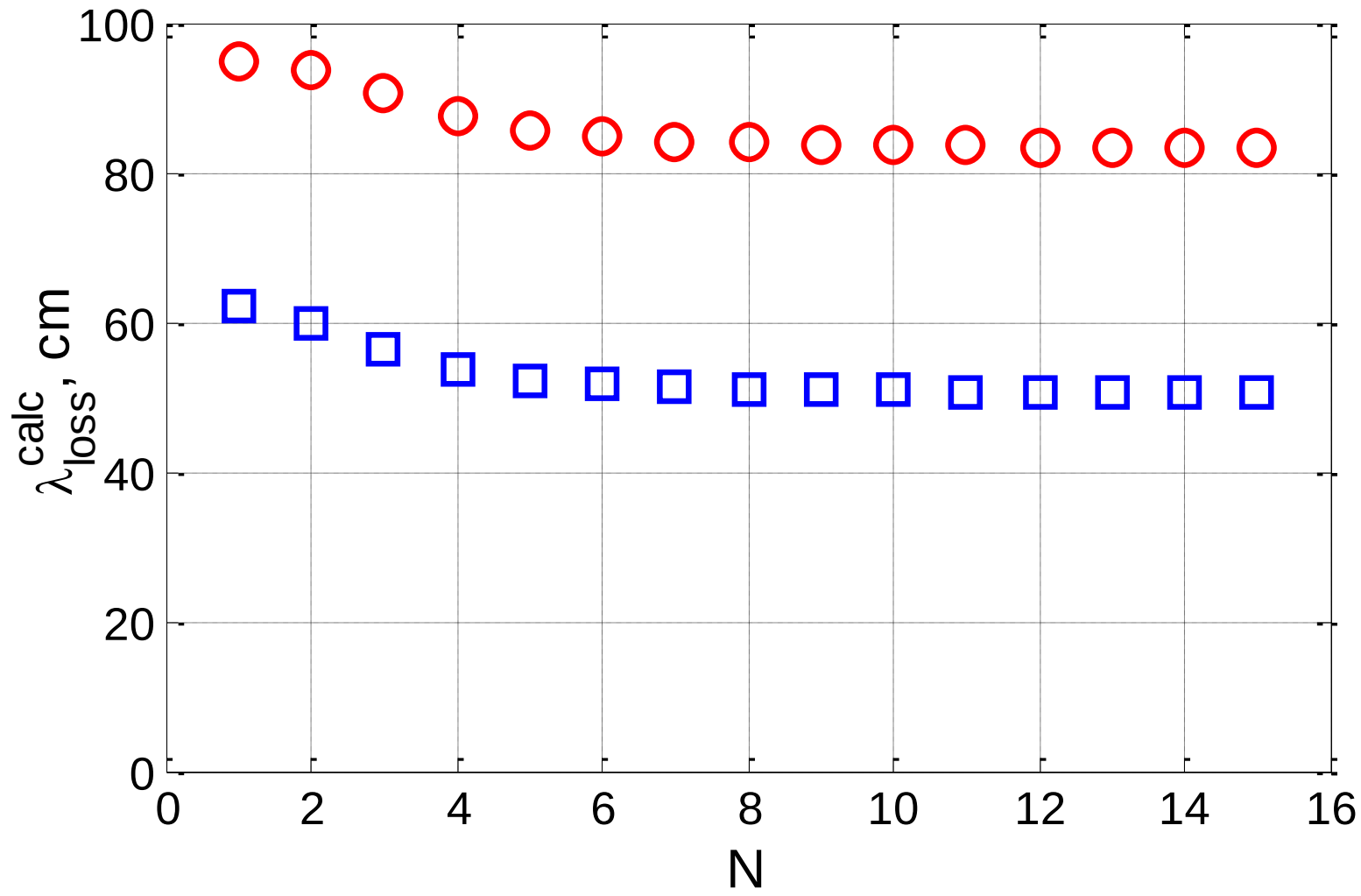

Figure 5. Dependence of the calculated mean free path of fast atoms at the injection energy on the number of levels included in the model; blue squares – GDML, red circles – GOL-NB.

The mean free path of fast atoms in the GDML facility for different beam components (with energies of 2.22, 13.3, 20, and 40 keV) is shown in Figure 6 (blue circles). For all energies, the contribution of collisions with thermal ions to the ionization of beam atoms is about 50%, while the contributions of collisions with electrons and fast ions are approximately 25% each. At the same time, due to stepwise ionization, there is a mutual influence of collisions with different plasma particles, such that the inverse mean free path of an atom in the plasma is not equal to the sum of the inverse mean free paths calculated considering collisions with each particle species (electrons, warm ions, and fast ions) separately. In Figure 6, red crosses show the mean free paths calculated under the assumption of additivity of the contributions from collisions with each plasma species. The interplay between different particle species leads to a reduction in the mean free path. This effect is more pronounced at high hydrogen atom energies; at an energy of 40 keV, the difference between the mean free paths calculated with a correct treatment of all particles and those obtained by summing the contributions of individual species without accounting for their mutual influence amounts to 7%.

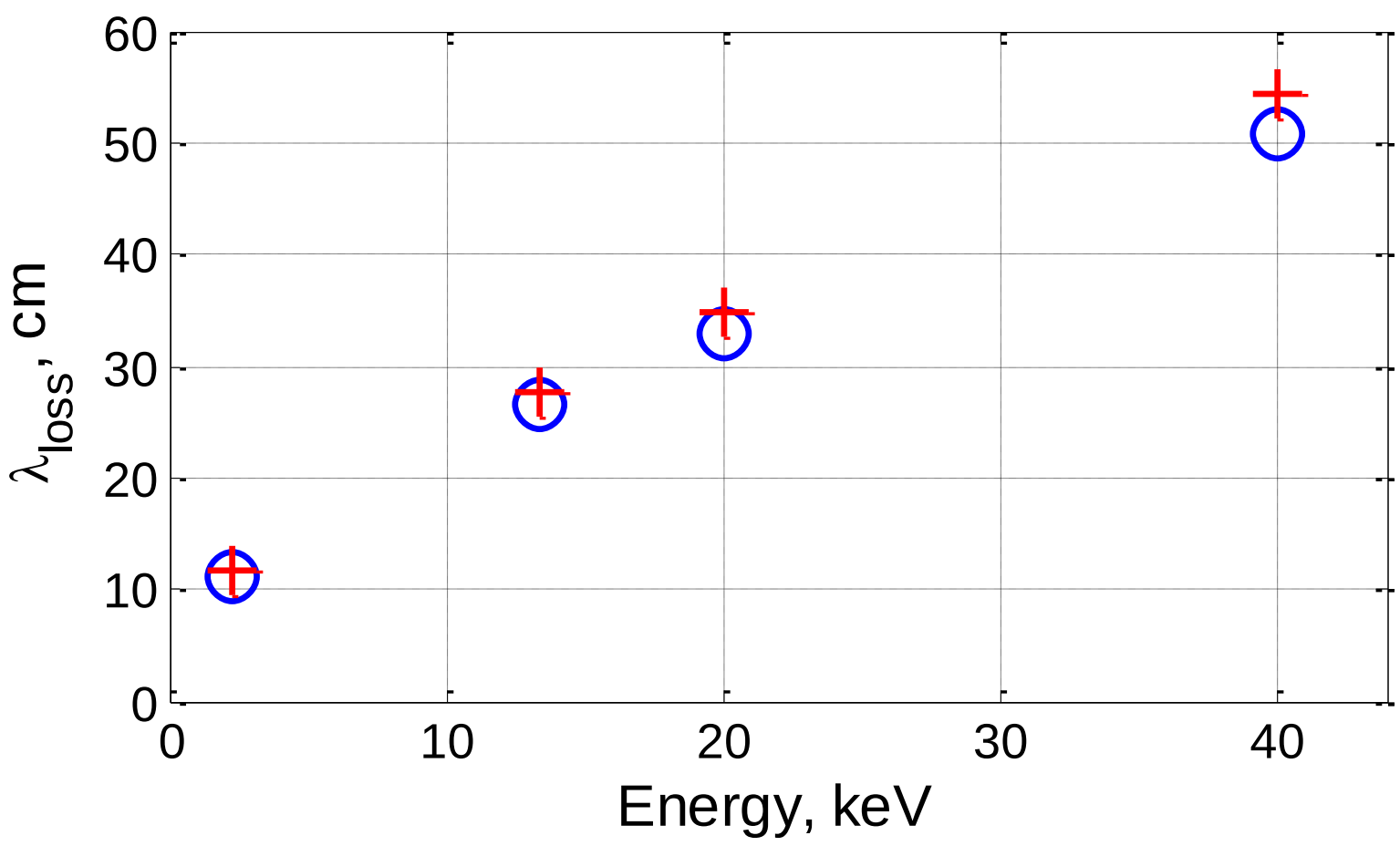

Fig. 6. Mean free path of fast hydrogen atoms in the GDML facility. Blue circles – exact calculations with simultaneous consideration of processes involving all plasma particles (electrons, warm ions, and fast ions); red crosses – mean free path obtained by summing the mean free paths calculated under the assumption of additivity of the contributions from collisions with each plasma species

Additionally, the effect of the anisotropy of the fast ion distribution function in the plasma was investigated. The HIR package contains a function for calculating the rates of elementary processes in collisions of fast atoms with plasma particles having a two-dimensional axisymmetric distribution function, as well as a similar function for collisions with particles having an isotropic velocity distribution. To verify the effect of anisotropy, calculations of the mean free path of fast atoms in the GDML facility were performed for the two-dimensional distribution function of fast ions (Fig. 4b) and also for a spherically symmetric distribution function with the same absolute velocity distribution (Fig. 4a). The calculations showed that the resulting mean free paths differ by less than 1%, meaning that for the GDML facility, it is permissible to use an isotropic distribution function in calculations of processes involving hydrogen atoms.

In the GOL-NB facility, the plasma consists only of thermal electrons and ions; therefore, the mean free paths of fast hydrogen atoms are conveniently characterized by the effective capture cross section per electron–ion pair. The effective capture cross sections as a function of plasma density at a temperature of $T_e$ = 20 eV for the full (24 keV) and fractional (12 and 8.3 keV) energies of fast atoms are shown in Figure 7 (solid blue curves). Capture occurs predominantly via collisions with plasma ions; the contribution of electron collisions to the effective cross section is about 20%. Due to stepwise ionization of hydrogen atoms, the effective capture cross section exhibits a dependence on plasma density; in the density range characteristic of the GOL-NB facility $10^{12}$– $10^{14}$ cm$^{-3}$, this cross section varies by 10%. The dashed curves in Figure 7 show the sum of the effective capture cross sections for collisions with electrons and ions separately. The non-additivity of the effective cross sections (also related to stepwise ionization) increases with increasing density and amounts to about 4% at $10^{14}$ $cm^{-3}$.

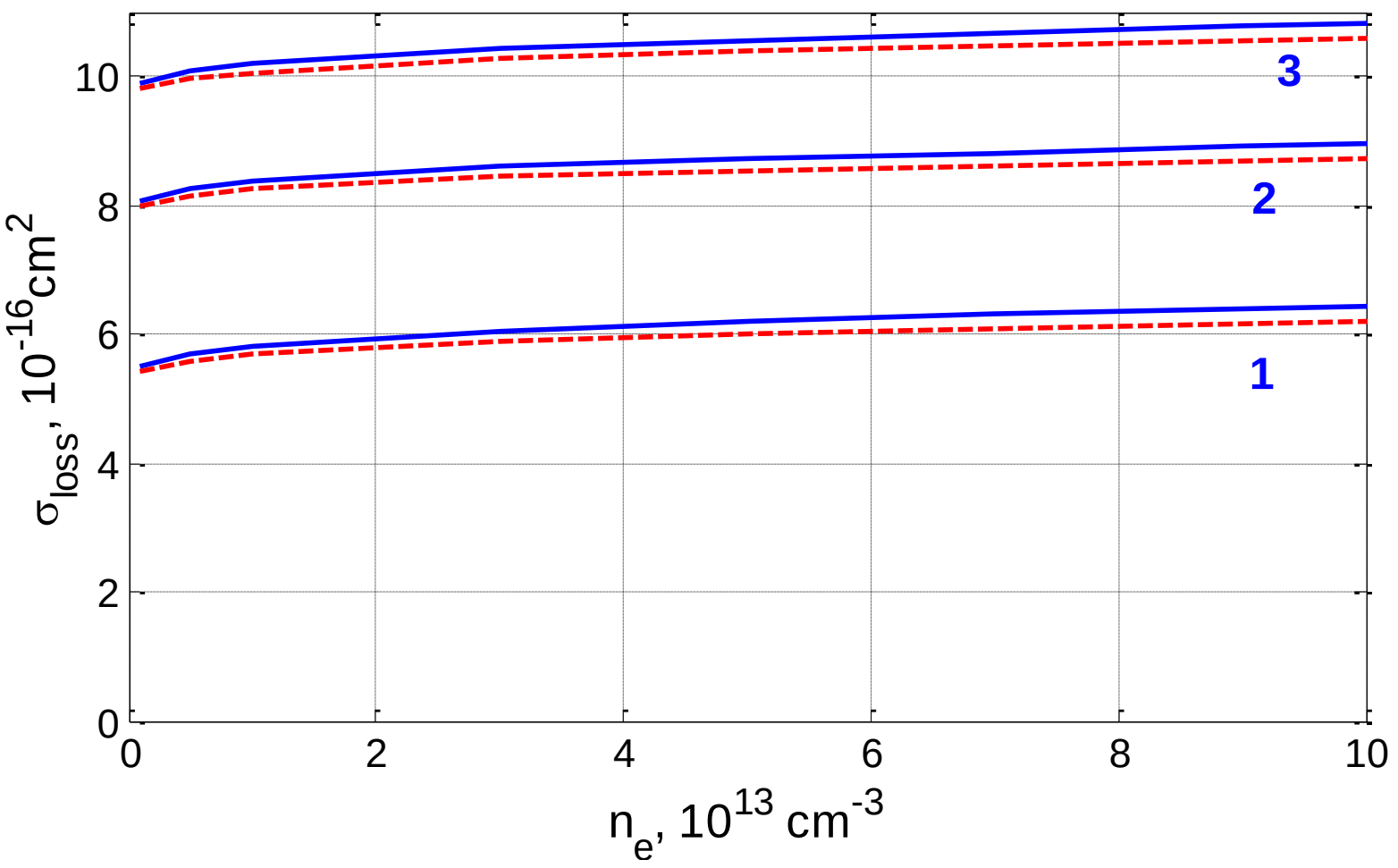

Fig. 7. Effective capture cross section of atomic beams in the plasma of the GOL-NB facility at $T_e$=20 eV. 1) – atom energy 24 keV; 2) – atom energy 12 keV; 3) – atom energy 8.3 keV. Solid blue curves – calculations of level populations with simultaneous consideration of collisions with plasma electrons and ions; red dashed curves – sum of the effective capture cross sections for collisions with electrons and ions separately.

## 5. Ionization cost

Due to the relatively low electron temperature in the GOL-NB facility, the line radiation of atoms entering the plasma to compensate for particle losses can play a significant role in the plasma energy balance. The energy costs associated with plasma fueling can be characterized by the ionization cost — the sum of the hydrogen atom ionization energy and the energy of photons emitted by the atom before its ionization in the plasma. The radiation energy losses per atom entering and ionizing in the plasma can be obtained from the level population distribution as:

$$Er = \frac{\sum_{k=2}^{n_{max}} n_k \left(\sum_{i=1}^{k-1} A_{ki}\, \Delta E_{ik}\right)}{\sum_i \Gamma_i}$$

The dependence of the ionization cost for atomic hydrogen on plasma temperature for different densities ($10^{12}$, $10^{13}$, and $10^{14}$ $cm^{-3}$) is shown in Figure 8. In the calculations of level populations, only collisions with electrons (ionization, excitation, and de-excitation) were taken into account, assuming that the secondary atom resulting from charge exchange remains in the plasma. It can be seen that as the plasma density increases, the ionization cost decreases, which is associated with the competition between collisional and radiative de-excitation of atoms.

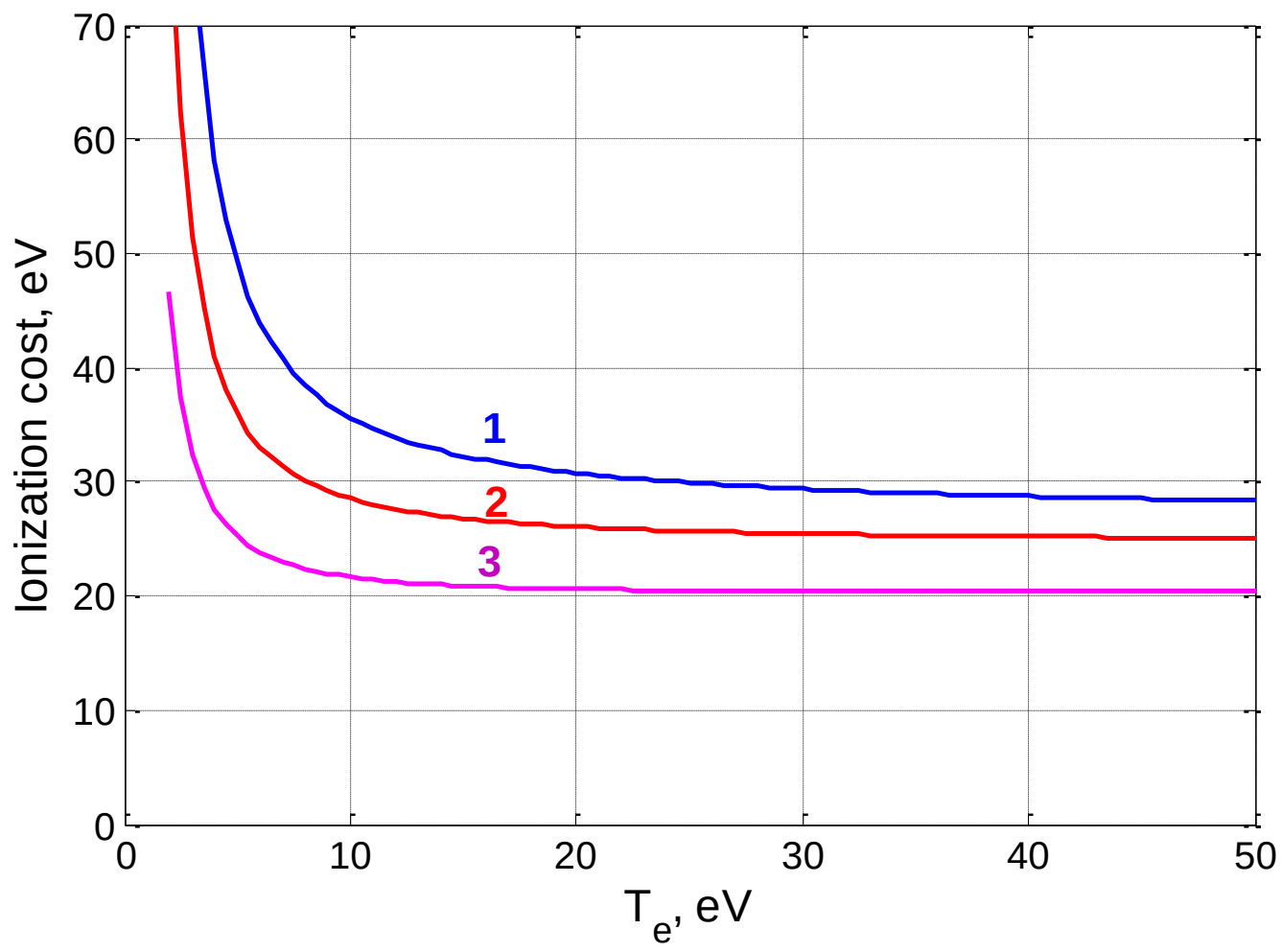

Figure 8. Ionization cost of a hydrogen atom in plasma as a function of plasma temperature for different densities; 1) – $n_e=10^{12}$ cm$^{-3}$, 2) – $n_e=10^{13}$ cm$^{-3}$, 3) – $n_e=10^{14}$ cm$^{-3}$.

## Conclusion

The developed software package HIR (Hydrogen Ionization Rates) allows one to calculate the level populations of atomic hydrogen in plasma and the dynamics of its ionization, including in the presence of fast ions with an anisotropic distribution function. Stepwise ionization of hydrogen atoms can make a noticeable contribution (on the order of 20%) to the ionization rate, and for a correct description of stepwise ionization, it is necessary to account for level populations up to n = 8. The effects of stepwise ionization lead to a dependence of the ionization rate and the effective capture cross section of fast atoms on plasma density, as well as to non-additivity of processes involving different plasma components.

The HIR package is publicly available and can be used for modeling neutral beam physics in both mirror and toroidal magnetic confinement systems

## Acknowledgements

The work is supported by Russian Science Foundation, grant No. 25-22-00117

# Supplement: HIR code description

## Physics basis

HIR (Hydrogen Interaction Rates) is a code for the simulation of processes involving atomic hydrogen in hot plasma with fast ions.
The code calculates hydrogen level populations using an equilibrium collisional-radiative model. The processes included are:

- electron impact excitation and de-excitation
- electron impact ionization
- ion impact excitation and de-excitation
- ion impact ionization
- ion-atom charge exchange
- spontaneous emission

Population equilibrium is achieved by balancing the influx of atoms into the plasma against their loss due to ionization and charge exchange. Recombination processes are not included in the model.

The system of equations solved in the code for determining the level populations of the atom is:

$$\sum_{k=i+1}^{n_{max}} n_k A_{ik} - \sum_{k=1}^{i-1} n_i A_{ki} - n_i(\nu_i^{ion_e} + \nu_i^{ion_i} + \nu_i^{CX}) + \sum_{k=i+1}^{n_{max}} n_k\left(\nu_{ik}^{deex_e} + \nu_{ik}^{deex_i}\right) - \sum_{k=1}^{i-1} n_i\left(\nu_{ki}^{deex_e} + \nu_{ki}^{deex_i}\right) = \Gamma_i$$

Here i,k – are the principal quantum numbers of the initial and final levels of the transitions, $\nu = n_{pl}\langle\sigma v\rangle$ is the collision rate of a specific process, $n_{pl}$ is the density of plasma particles colliding with the atom in that process.
$\Gamma_i$ is the incoming flux of atoms with an electron on level i. For incoming atoms in the ground state $\Gamma_1=1$, $\Gamma_{i>1}=0$.
$A_{ik}$ are the spontaneous radiation transition probabilities.
The atomic level population $n_i$ can be used to calculate quantities characterizing the interaction of the atom with different plasma components..
For example, the global lifetime of atoms prior to their conversion to ions (due to ionization or charge exchange) is:

$$\tau = \frac{\sum_i n_i}{\sum_i \Gamma_i}$$

The partial lifetime prior to a specific process is:

$$\tau_p = \frac{\sum_i n_i}{\sum_i n_i \cdot \nu_i^{spec}}$$

The effective ionization rate is

$$\langle\sigma v\rangle_{eff} = \frac{\sum_i n_i \cdot (\nu_i^{ion_e} + \nu_i^{ion_i} + \nu_i^{CX})}{n_{pl}\sum_i n_i}$$

The energy loss due to spontaneous emission, normalized to the number of captured atoms, is:

$$Er = \frac{\sum_{k=2}^{n_{max}} n_k\left(\sum_{i=1}^{k-1} A_{ki} E_{ki}\right)}{\sum_i \Gamma_i}$$

## Processes cross section and reaction rates

The code use the data for cross section of interaction of electrons and ions with hydrogen atom from the report [R.K. Janev and J.J. Smith Cross sections for collision processes of hydrogen atoms with electrons, protons and multiply charged ions, Atomic And Plasma-Material Interaction Data For Fusion, V.4 AEA-APID-4 (1993)].

### Code structure

The code uses Python 3.8.5 with the following libraries:
NumPy – for numerical computations and array operations
SciPy – (optional) for loading MAT-files of user-defined distribution function
Numba – (optional) for Just-In-Time (JIT) compilation to accelerate calculations
dataclasses – for defining parameter containers
ABC (Abstract Base Classes) – for creating abstract class hierarchies"

The code incudes the follows components:
Basic code components:

**Cross_section_func.py, Cross_section_func_jit.py** – Functions for calculation cross-section of atomic processes
**Distribution_Functions.py** – Classes of plasma particles distribution functions
**Atomic_Processes.py** – Classes of atomic processes
**set_calc_parameters.py** – Functions for setting calculation parameters
**HIR_Util.py**– utilities for calculations and post-processing

Examples of calculations:

**Radiation_power.py** – calculation of energy radiated by hydrogen atom before ionization in the plasma of GOL-NB facility vs. plasma density
**Atom_path_length_density_scan** – calculation of path length of fast atom in plasma of GOL-NB facility vs. plasma density
**Atom_path_length_ energy_scan** – calculation of path length of fast atom in plasma of GDMT facility for beam components with different energy

### *Cross_section_func.py, Cross_section_func_jit.py*

Approximation formulas for cross sections of hydrogen excitation and ionization by electron and proton impact, as well as proton charge exchange, are taken from [*R.K. Janev and J.J. Smith, "Cross sections for collision processes of hydrogen atoms with electrons, protons and multiply charged ions," Atomic and Plasma-Material Interaction Data for Fusion, Vol. 4, AEA-APID-4, 1993*]. De-excitation is calculated using the Klein–Rosseland formula:

$$\sigma_{deex}(E) = \sigma_{ex}(E + \Delta E_{ik}) \cdot \frac{g_i}{g_k} \cdot \frac{E + \Delta E_{ik}}{E}$$

Both excitation and de-excitation are calculated in the same function: ***h_e_ex_deex(v_in, n, m, process)*** for electron impact and ***h_p_ex_deex(v_in, n, m, process)*** for proton impact, where process = 1 corresponds to excitation and process = 0 to de-excitation.

The spontaneous radiation transition probability is calculated using the Johnson formula [*L.C. Johnson, The Astrophysical Journal, 174:227–236, 1972*].

There are two versions of the functions: one written in pure Python (***Cross_section_func.py***) and one using JIT compilation (***Cross_section_func_jit.py***). The version to be used is determined by the JIT property of the instance of CalcParameters class in ***HIR_Util.py*** (see below).

### *Distribution_Functions.py*

The classes for defining distribution function of plasma particles and calculating of specific process rates <σv>.
The classes are

- **ElectronDF** - Maxwellian DF of electrons with $T_e$ defined in an instance of the class CalcParameters
- **WarmIonDF** - Maxwellian DF of ions with $T_i$ defined in an instance of the class `CalcParameters`
- **FastIonDF** - Isotropic DF of ions imported from an external file (file name defined in an instance of the class `CalcParameters`)
- **AnisotropicFastIonDF** - 2D (with axial symmetry) anisotropic DF of ions imported from an external file (file name defined in an instance of the class `CalcParameters`)

The methods of the classes are:

- **define_DF()**
  Produces a structured NumPy array with the fields **'velocity'**, **'Energy'**, **'distr'** for isotropic DF or **'v_parallel'**, **'v_transverse'**, **'Energy'**, **'distr'** for 2D anisotropic DF

- **recalc_DF()**
  Re-calculate distribution function if a parameter of DF (temperature in `CalcParameters`) has changed. Executed in every rate calculation.

- **rate_calculation(sigma_fun, sigma_kwargs)** or **rate_calculation_jit(self, sigma_fun, n, m, process)**
  Calculate process rates (<σv>) using the input cross-section function `sigma_fun` for defined distribution function of plasma particles and fast (non-zero velocity) atom. Atom velocity and angle relative to the symmetry axis are defined in instances of the class `CalcParameters`.
  There are two versions of the methods – pure and JIT. The version to be used is determined by the `JIT` property of the instance of `CalcParameters` class in ***HIR_Util.py***.

  2D anisotropic distribution function should be defined such as

$$\frac{1}{2}\cdot\iint f(v_\perp, v_\parallel)\cdot 2\pi\cdot v_\perp\cdot dv_\perp dv_\parallel = 1$$
$$0 < v_\perp < \infty$$
$$0 < v_\parallel < \infty$$

### *Atomic_Processes.py*

The classes for calculating transition frequency matrix for different atomic processes are:

- `ElectronImpactExcitation` - electron excitation
- `IonImpactExcitation` - ion excitation
- `ElectronImpactDeexcitation` - electron de-excitation
- `IonImpactDeexcitation` - ion de-excitation
- `ElectronImpactIonization` - ionization by electron impact
- `IonImpactIonization` - ionization by ion impact
- `IonChargeExchange` - charge exchange with ions
- `Emission` - spontaneous radiative transitions

The main method of the classes is `calculate_matrix()`. This method calls the method `rate_calculation` of a `DitributionFunctons` class instance for every upper and lower level n and m, using the specific cross-section function for the corresponding atomic process. The calculated rates are then assembled into transition frequency matrix ν.

### *set_calc_parameters.py*

The functions perform the following operations: they construct a `CalcParameters` instance and assign values to its required properties; create `DistributionFunctions` objects (representing distribution functions of different plasma species) based on this instance; and instantiate `Atomic_Processes` objects for all combinations of atomic processes and impact-particle distribution functions.
Two examples of these functions are:
`set_calc_parameters_GDMT()` – for the GDMT facility
`set_calc_parameters_GOL_NB()` – for the GOL-NB facility

### *HIR_Util.py*

Utilities for calculation:
class `Const` – a static class with class variables containing the general physical constants used in calculations.

class `CalcParameters`:
The class for storing and transmitting calculation parameters to all used functions and methods. The class includes two methods:

`calc_v_shift()` – calculates the atom velocity from its energy
`calc_ultimate_level()` – calculates the ultimate level of a fast hydrogen atom in a magnetic field (due to autoionization caused by the motional Stark electric field):

$$N = \frac{1}{2}\left(\frac{E_B}{\left|\vec{v} \times \vec{B}\right|}\right)^{1/4} = \frac{1}{2}\left(\frac{Ry}{e \cdot a_B} \cdot \frac{1}{\left|\vec{v} \times \vec{B}\right|}\right)^{1/4}$$

`find_population_matrix(CalcPar, process_list)` – the main function for constructing the transition frequency matrix and solving the system of balance equations to determine the atomic level populations.

Post-processing functions:
`calc_path_length(CalcPar, all_processes)` – calculates the path length of a fast atom in the plasma until capture (ionization or charge exchange)
`calc_radiation_losses(CalcPar,all_processes)` - calculates the energy radiated until the hydrogen atom is ionized in the plasma (eV/atom)